\documentclass{agujournal2019}
\usepackage{url}
\usepackage{soul}

\draftfalse

\journalname{JGR: Planets}

\usepackage{amsmath,amssymb,amsfonts,amsthm}
\usepackage[version=4]{mhchem}
\usepackage[separate-uncertainty=true, multi-part-units=single,  range-units=single, retain-explicit-plus=true]{siunitx}

\newcommand{\figref}[1]{Figure~\ref{#1}}
\newcommand{\tabref}[1]{Table~\ref{#1}}
\newcommand{\secref}[1]{Section~\ref{#1}}

\let\citet\citeA

\begin{document}

\title{Global Modelling of Ganymede's Surface Composition: Near-IR Mapping from VLT/SPHERE}
\authors{Oliver King\affil{1}, Leigh N Fletcher\affil{1}}
\affiliation{1}{School of Physics and Astronomy, University of Leicester, University Road, Leicester, LE1 7RH, United Kingdom}
\correspondingauthor{Oliver King}{ortk1@le.ac.uk}

\begin{keypoints}
	\item VLT/SPHERE maps show Ganymede's surface is mainly water ice in young terrain and a dark, spectrally flat material in old terrain.
	\item Water ice grain size varies on global scales, controlled by radiation driven sputtering and temperature gradients.
	\item Hydrated salts and sulphuric acid have lower abundances, with uncertainties making it difficult to make confident detections.
\end{keypoints}

\begin{abstract}
	We present maps of Ganymede's surface composition with almost complete longitude coverage, acquired using high spatial resolution near-infrared (\SIrange{0.95}{1.65}{\micro\m}) observations from the ground-based VLT/SPHERE instrument. Observed reflectance spectra were modelled using a Markov Chain Monte Carlo method to estimate abundances and associated uncertainties of water ices, acids, salts and a spectrally-flat darkening agent. Results confirm Ganymede's surface is dominated by water ice in young bright terrain (impact craters, sulci), and low-albedo spectrally-flat material in older dark terrain (e.g. Galileo Regio). Ice grain size has strong latitudinal and longitudinal gradients, with larger grains at the equator and on the trailing hemisphere. These trends are consistent with the effects of the latitudinal thermal gradient and global variations in radiation driven sputtering. Sulphuric acid has a low abundance and appears potentially spatially correlated with plasma bombardment, where Ganymede's poles are exposed to the external jovian magnetic field. Best-estimate abundances suggest a mixture of salts could be present, although their low abundances, spectral degeneracies and associated uncertainties mean individual salt species cannot be detected with confidence. If present, sodium magnesium sulphate and magnesium chlorate appear tentatively correlated with exogenic plasma bombardment, while magnesium chloride and sulphate appear tentatively correlated with younger terrain, implying a possible endogenic origin. MCMC modelling was also performed on Galileo/NIMS data, showing comparable distributions. The high spatial resolution of SPHERE allows the precise mapping of small scale (\SI{<150}{km}) surface features, which could be used along with higher spectral resolution observations to jointly confirm the presence and distribution of potential species.
\end{abstract}

\section*{Plain Language Summary}
We have observed Jupiter's icy moon Ganymede, the solar system's largest moon, with the Very Large Telescope in Chile. These observations recorded the amount sunlight reflected from Ganymede's surface at different infrared wavelengths, producing a `reflectance spectrum'. We used these reflectance spectra to understand the surface composition of Ganymede by using a computer model that compares each observed spectrum to spectra of different substances that have been measured in laboratories. Ultimately, this model allows us to calculate the amount of each material at each location on Ganymede's surface.

Our results show how Ganymede's surface is made up to two main types of terrain: young areas have large amounts of water ice, whereas ancient areas mainly consist of a dark grey material which we were unable to identify. We detected sulphuric acid near Ganymede's poles, which is likely to originate from the gasses which surround Jupiter. A range of different salts were detected, some of which may originate from within Ganymede itself. These surface composition maps will be useful for understanding the processes happening on, and under, Ganymede's surface. They will also help to plan for the robotic space missions which are due to explore Ganymede up-close in the coming decades.

\section{Introduction} \label{sec:intro}
Ganymede, the largest of Jupiter's Galilean moons, is the largest moon in the solar system, with a radius of \SI{2631}{km}. It has a strongly differentiated interior and intrinsic magnetic field, likely caused by convection in its iron core \cite{kivelson1996discovery,schubert2004interior}. Ganymede is believed to have a subsurface ocean, likely sandwiched between multiple layers of ice \cite{kivelson2002permanent}. Ganymede's crust is mainly composed of water ice with significant contamination from non-ice materials on the surface \cite{pappalardo2004geology, schubert2004interior,mccord1998non,johnson1983global}.

Much of the current understanding of Ganymede's surface composition comes from the Galileo mission that orbited Jupiter from 1995 to 2003 with repeated flybys of the Galilean satellites. Observations with the Near-Infrared Mapping Spectrometer (NIMS) \cite{carlson1992near} and panchromatic cameras identified a surface made up of contrasting dark and light terrain, covering about 1/3 and 2/3 of the surface, respectively. The dark terrain (`regiones') is heavily cratered, suggesting it is \SI{>4}{Gyr} old and appears to be covered with a thin layer of spectrally-flat dark material \cite{pappalardo2004geology,schenk2004ages}. The light terrain (`sulci') is much younger, with fewer craters and cleaner exposed ice, suggesting the light regions were caused by tectonic events, potentially through rift like formation or cryovolcanic resurfacing \cite{pappalardo2004geology,schenk2004ages}.

Ganymede's intrinsic magnetic field was discovered by the Galileo mission \cite{kivelson1996discovery} and leads to complicated interactions between the surface and Jupiter's magnetospheric plasma environment. Equatorial latitudes ($\lesssim \ang{40}$) have closed magnetic field lines, so are protected from significant exogenic plasma bombardment. Higher latitudes, however, have open field magnetic lines, exposing Ganymede's polar regions to plasma bombardment from Jupiter's magnetosphere \cite{liuzzo2020variability}. This bombardment delivers ions that can then be radiolytically converted on Ganymede's surface into detectable species. For example, sulphur ions from Io's volcanism have been identified as the cause of high sulphuric acid concentrations on Europa's trailing hemisphere where plasma bombardment is highest \cite{carlson2005distribution, brown2013salts, ligier2016vlt, king2022compositional}. Magnetospheric modelling suggests that the polar regions on Ganymede's leading hemisphere are likely to experience the highest rates of plasma bombardment, and therefore are likely to have the highest abundance of exogenic material \cite{liuzzo2020variability}, such as the sulphuric acid detected by \citet{ligier2019surface}.

Ganymede has distinct polar caps, consisting of water frost coating the surface at latitudes $\gtrsim\ang{40}$ \cite{mccord2001hydrated}. Analysis of Galileo observations and modelling of Ganymede's intrinsic magnetic field \cite{khurana2007origin} show that the edge of the polar cap is well correlated with the edge of the icy polar cap. This suggests the polar cap is likely to be caused by sputtering of water ice (due to the higher plasma bombardment at high latitudes) redistributing water ice which is then trapped on the surface at the cold polar temperatures \cite{khurana2007origin}. The thermal gradient between Ganymede's equator and poles also causes a gradient in ice grain sizes which has been observed with NIMS \cite{mccord2001hydrated,stephan2020h2o}.

More recent advances in telescope optics have enabled studies using ground-based observatories to map compositional contrasts on Ganymede. \citet{ligier2019surface} used VLT/SINFONI (\SIrange{1.40}{2.50}{\micro m}, $R\approx2000$, \SI{\sim70}{km/px}) to find similar results to Galileo/NIMS with a surface dominated by water ice and spectrally-flat darkening agent, with smaller amounts of hydrated sulphuric acids at high latitudes and a relatively uniform low distribution of potentially endogenic hydrated salts.

Studies have been unable to identify the composition of the darkening agent that dominates Ganymede's older terrain. Hydrated silicates, carbonaceous compounds, or potentially some hydrated salts, are the most likely candidates, but the lack of unique spectral features makes any determination very difficult \cite{pappalardo2004geology,ligier2019surface,stephan2020h2o}. The exact mixture of hydrated salts on Ganymede's surface is also uncertain, with a range of sulphates and chlorinated salts all candidates \cite{ligier2019surface,mccord1998non}. Many of these salts have similar spectral features, generally distorting water ice bands \cite{mccord2001hydrated}, so a range of salt mixtures appear consistent with observed Ganymede spectra.

The Juno orbital mission of Jupiter has enabled high spatial resolution observations of Ganymede with the JIRAM spectrometer (\SIrange{2}{5}{\micro\m}). Observations of Ganymede's northern hemisphere \cite{mura2020infrared} identified higher water ice abundance in Ganymede's polar cap (\ang{>45}N) and various absorption features which may be caused by hydrated salts. Continuing observations through the Juno mission will allow more detailed study of Ganymede's surface composition, especially with the unprecedented high spatial resolutions (\SI{<1}{km/px}) offered by Juno's position within the Jupiter system \cite{tosi2021ganymede}.

In this paper, we will discuss our analysis of observations of Ganymede's near-infrared reflected sunlight spectrum using the SPHERE instrument on the ground-based Very Large Telescope, VLT \cite{beuzit2019sphere}. These observations will be used to analyse the composition of the Ganymede's surface, in relation to the physical and chemical processes shaping icy worlds. The VLT/SPHERE (\SI{\sim25}{km/px}) observations provide significant improvements in spatial resolution compared previous VLT/SINFONI (\SI{\sim75}{km/px}) ground-based observations of Ganymede \cite{ligier2019surface}, although SPHERE has a lower spectral resolution than SINFONI. We will also compare the SPHERE results to Galileo/NIMS observations to demonstrate that ground-based observations are capable of reproducing and extending those from visiting spacecraft, enabling them to provide key support for future missions.

\section{Observations and data reduction} \label{sec:reduction}
\subsection{VLT/SPHERE}
\begin{table}
	\caption{VLT/SPHERE observation log. \label{tab:observation-log}}
	\centering
	\begin{tabular}{llrr}
		\hline
		\textbf{Time (UTC)}    & \textbf{Target} & \textbf{Sub-observer point} & \textbf{Phase angle} \\
		\hline
		04:19 1 February 2015  & Ganymede        & \ang{116.9}W,~\ang{0.1}S    & \ang{1.215}          \\
		04:01 21 July 2021     & Ganymede        & \ang{25.6}E,~\ang{0.7}N     & \ang{6.154}          \\
		04:34 21 July 2021     & HD 213199       &                             &                      \\
		05:00 22 July 2021     & Ganymede        & \ang{26.8}W,~\ang{0.7}N     & \ang{5.981}          \\
		05:38 22 July 2021     & HD 213199       &                             &                      \\
		04:53 5 September 2021 & Ganymede        & \ang{136.3}W,~\ang{0.6}N    & \ang{3.490}          \\
		05:18 5 September 2021 & HD 213199       &                             &                      \\		\hline
	\end{tabular}
\end{table}

\begin{figure}
	\centering
	\includegraphics[width=0.667\linewidth]{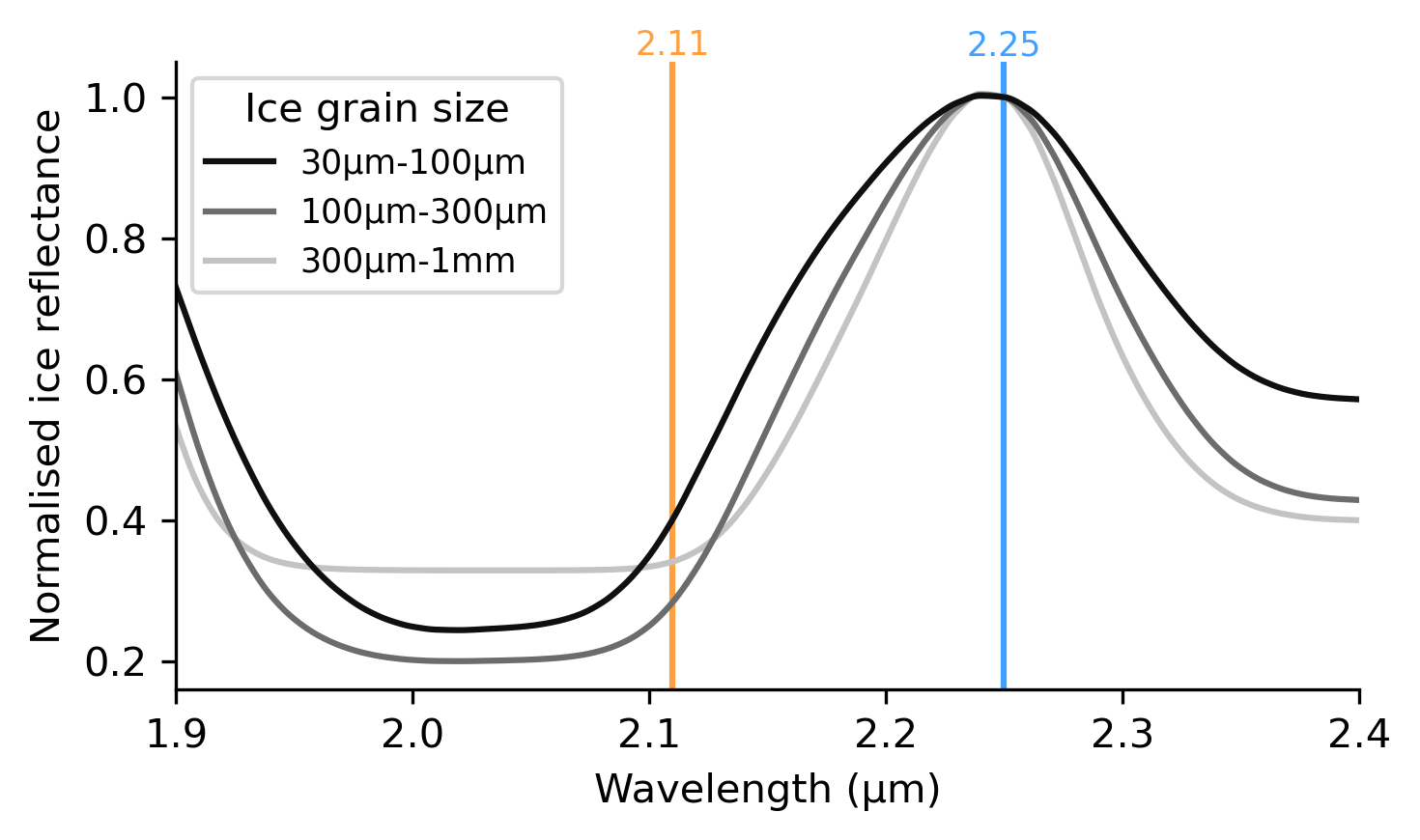}
	\caption{\SI{2}{\micro\m} water ice absorption band for a range of water ice grain sizes. Spectra are normalised to unity at \SI{2.25}{\micro\m} to show the relative band depths; full un-normalised spectra showing the absolute band depths are shown in \figref{fig:example-endmembers}. The central wavelengths of the two IRDIS filters are shown by the vertical lines.
		\label{fig:irdis-ref-spec}}
\end{figure}
\begin{figure*}
	\centering
	\includegraphics[width=\linewidth]{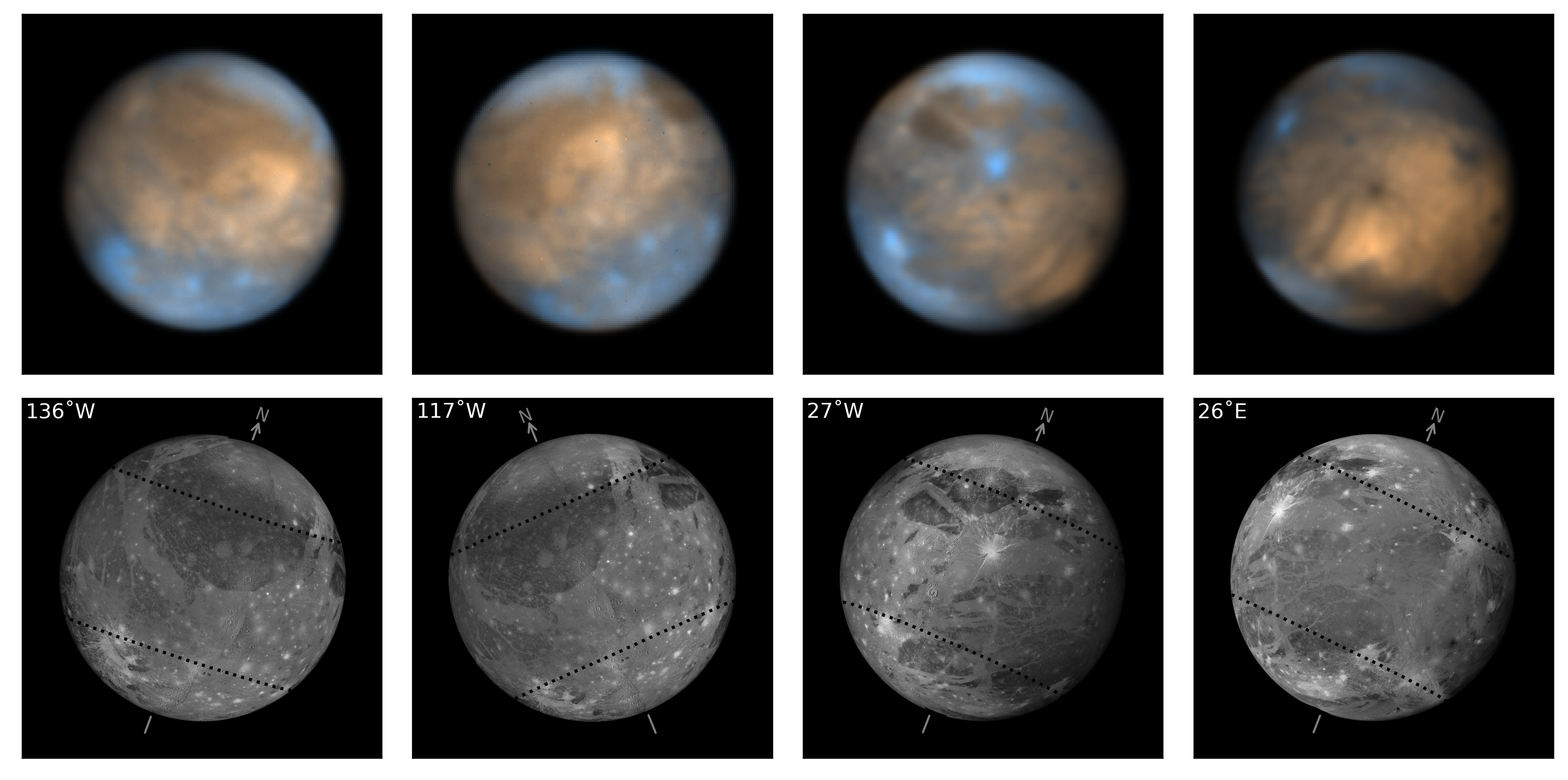}
	\caption{IRDIS dual band images of Ganymede (top) compared to simulated visible light images (bottom). The two-colour IRDIS images are produced using the K1 filter at \SI{2.11}{\micro\m}, shown in orange, and the K2 filter at \SI{2.251}{\micro\m} in blue. Water ice has a broad absorption band around \SI{2}{\micro\m} (see \figref{fig:irdis-ref-spec}), so the blue areas are icy and the orange areas are contaminated with non-ice material. The simulated images (bottom) use the visible light mosaic of Ganymede from \citet{usgs2013ganymede}; dotted black lines indicate Ganymede's approximate open/closed magnetic field line boundaries from \citet{khurana2007origin}.
		\label{fig:irdis-two-colour}}
\end{figure*}

The SPHERE (Spectro-Polarimetric High-contrast Exoplanet REsearch) instrument \cite{beuzit2019sphere} is located at the European Southern Observatory's Paranal Observatory on the Very Large Telescope (VLT). Originally designed to directly image exoplanets, VLT/SPHERE uses adaptive optics to provide high spatial resolution imaging. This resolution can be applied to solar system targets, enabling high spatial resolution mapping of objects such as Europa \cite{king2022compositional} and Ganymede.

Observations of Ganymede were taken during an observing campaign in 2021 and during science verification in 2015, as summarised in \tabref{tab:observation-log}. We used SPHERE in \verb|IRDIFS_EXT| mode, allowing simultaneous imaging with the Integral Field Spectrograph (IFS) and Infrared Differential Imaging Spectrometer (IRDIS) sub-systems of the SPHERE instrument. This allows simultaneous spectroscopic observations covering \SIrange{0.95}{1.65}{\micro\m} and dual band imaging at \SI{2.11}{\micro\m} and \SI{2.251}{\micro\m}.

The IFS \cite{claudi2008sphere, mesa2015performance} simultaneously records spatial and spectral information on the same detector, ultimately producing image cubes with 38 wavelength channels from \SIrange{0.95}{1.65}{\micro\m}. It has a low spectral resolution of $R = \lambda/\Delta\lambda \sim 30$ but a high spatial resolution, with a pixel size of \SI{7.46}{mas/px}, corresponding to \SI{\sim25}{km/px} at Jupiter. Accounting for diffraction, this allows features \SI{<150}{km} across to be clearly resolved on Ganymede's surface \cite{king2022compositional}. This is a significant improvement in spatial resolution to the VLT/SINFONI global compositional maps of Ganymede \cite{ligier2019surface}, and is comparable to many Galileo/NIMS observations which have limited spatial coverage.

IRDIS \cite{dohlen2008infra, vigan2010photometric} produces simultaneous imaging through dual filters on two parts of the same detector. Our observations use the \verb|DB_K12| filter set on IRDIS which have transmissions centred at \SI{2.11}{\micro\m} for the K1 filter and \SI{2.251}{\micro\m} for the K2 filter, with filter widths of \SI{0.051}{\micro\m} and \SI{0.055}{\micro\m}, respectively. These filters were selected so that IRDIS is sensitive to the water ice absorption band at \SI{2}{\micro\m}, as shown in \figref{fig:irdis-ref-spec}. The IRDIS dual band images for each individual Ganymede observation are shown in \figref{fig:irdis-two-colour}.

Uncertainties on the observations were calculated by averaging the noise level in areas observing the background sky to allow the calculation of the signal to noise ratio. For the ground-based SPHERE observations, the sky background at the edge of the calibration star observations was used to calculate the noise level by treating areas more than \SI{45}{pixels} from the star as background. This reduces the influence of any stray light from atmospheric scattering of Ganymede's bright disk (or from the calibration star itself) affecting this noise calculation.

The data reduction followed the scheme developed for our VLT/SPHERE observations of Europa, as described fully in \citet{king2022compositional}. The data are reduced using the standard ESORex instrument pipeline with custom image cleaning code to remove bad pixels and striping artefacts introduced across the detector. Images are photometrically corrected using the Oren-Nayar photometric correction \cite{oren1994generalization} to remove brightness variations across this disk and account for the effect of different phase angles. The use of this photometric correction allows the use of data up to an emission angle of $e=\ang{75}$, corresponding to $\sim90\%$ of the observed disc \cite{king2022compositional}. The Oren-Nayar correction has previously been used to photometrically correct observations of Ganymede with VLT/SINFONI \cite{ligier2019surface}. The data are calibrated using calibration star observations to remove telluric contamination in the spectra. The calibration star was not successfully observed for the 2015 observation, so it was calibrated using the regions of Ganymede's surface which were observed by both the 2015 the 2021 observations. Once cleaned and photometrically corrected, the data are mapped onto an equirectangular grid to allow easy dataset comparison and referencing to known geological features.

As a final step of the calibration, the brightness and spectral slope of the SPHERE data is adjusted to be equal to the brightness and spectral slope of the Galileo/NIMS observation (\secref{sec:nims}) using the area of overlap of datasets, as in \citet{king2022compositional}. This helps to remove any residual errors from the calibration pipeline, such as slightly differing spectral slopes from the calibration star spectrum and the solar spectrum. This adjustment also removes the effect of the phase angle variation between the datasets (normalising them all to the \ang{29} of the NIMS dataset) which is particularly important for the SPHERE data, which is observed close to opposition, so would be affected by Ganymede's bright opposition surge if uncorrected \cite{hapke1993book}. The normalisation also helps to ensure that observational datasets have consistent photometric conditions to the spectral library, as laboratory measurements of endmember reflectance spectra are typically carried out at a non-zero phase angle to avoid the opposition surge \cite{dalton2012low}.

It is important to note that any residual photometric errors will affect the quality of spectral fitting and introduce additional uncertainties by producing a final dataset that is too bright or dark. For example, a photometric correction that produces a spuriously dark spectral cube may produce modelling results with a larger amount of darkening agent than is physically present to account for this residual photometric error. Future work may improve the quality of photometric correction (and therefore the quality of the overall dataset and results) by using more sophisticated photometric models \cite{hapke1993book}, potentially incorporating regionally variable photometric properties across Ganymede's surface, as described in \citet{belgacem2021regional}.

\subsection{Galileo/NIMS} \label{sec:nims}
\begin{figure}
	\centering
	\includegraphics[width=0.75\linewidth]{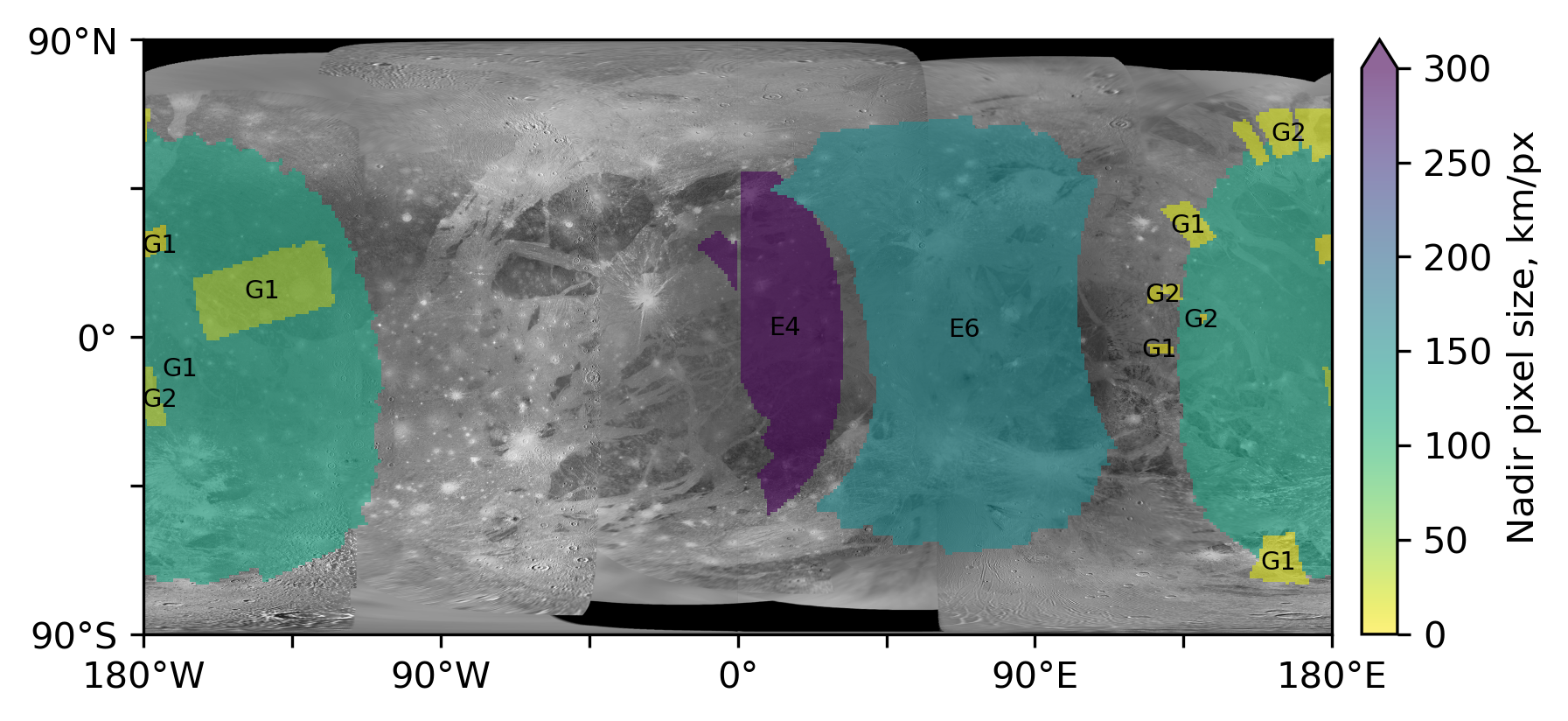}
	\caption{Coverage and spatial resolution of Galileo/NIMS observations with full coverage of SPHERE spectral range. Observations are labelled with the names of the Galileo orbits during which they were taken. Our SPHERE dataset has a spatial resolution \SI{<150}{km}.
		\label{fig:nims-resolution}}
\end{figure}

The Galileo mission orbited Jupiter from 1995 to 2003, during which a series of flybys of Ganymede enabled detailed imaging and mapping of Ganymede's surface. The Near-Infrared Mapping Spectrometer (NIMS) \cite{carlson1992near}, instrument on Galileo performed spectroscopy from \SIrange{0.7}{5.2}{\micro\m} with a spectral resolution of $R = \lambda/\Delta\lambda \sim 60$. During flybys, scans by the NIMS instrument produced spatially resolved infrared spectra in the region beneath the spacecraft. The flyby nature of the mission produced uneven spatial coverage, with some regions observed at very high spatial resolutions (\SI{\sim1}{km/px}) and others only observed at low spatial resolutions or completely unobserved (see \figref{fig:nims-resolution}).

The NIMS detector covering the \SIrange{0.99}{1.26}{\micro\m} wavelength range failed early in the Galileo mission, meaning that observations of Ganymede taken from 1997 onwards do not have full spectral coverage. Therefore, there are large areas of Ganymede's surface which lack complete NIMS observations (see \figref{fig:nims-resolution}). The SPHERE IFS wavelength range includes this missing spectral range so SPHERE can fill this gap in the spatially resolved near-infrared spectra of Ganymede.

In this study, we use the `G1GNGLOBAL01A' NIMS observation (taken at 1996-06-26 21:28 UTC at phase angle of \ang{29}) as a comparison for our VLT/SPHERE observation. This observation a similar spatial resolution (\SI{\sim 125}{km/px}) to our SPHERE dataset (\SI{<150}{km}), so can be used to provide a direct comparison of modelling results of the same region with both the ground-based SPHERE and space-based NIMS. This NIMS observation of Ganymede was downloaded from the NASA Planetary Data System, and reduced using a pipeline similar to that for our ground based observations \cite{king2022compositional}. The NIMS observations were photometrically corrected using the same Oren-Nayar photometric model as the SPHERE dataset, and show good consistency with the SPHERE dataset, especially towards the centre of the observation where the selected case study sites are located. The I/F NIMS data cubes were processed using ISIS3 \cite{isis} to calculate the viewing angles and coordinates that were then used as inputs for our mapping and photometric code.

\section{Water ice absorption bands}
\begin{figure*}
	\centering
	\includegraphics[width=\linewidth]{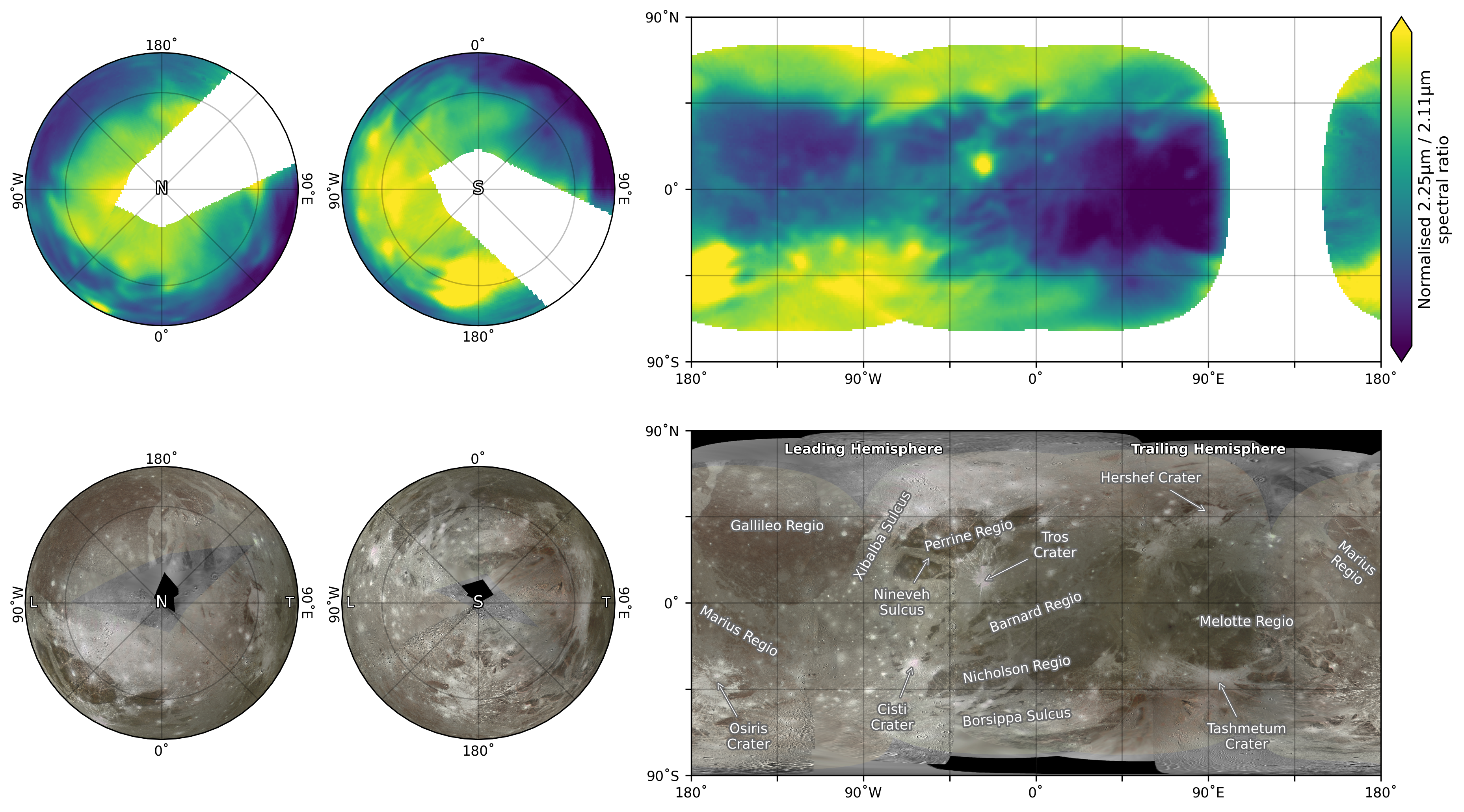}
	\caption{Normalised $2.25/\SI{2.11}{\micro\m}$ spectral ratio as measured by the dual band IRDIS instrument (top). Water ice has a strong absorption at \SI{2}{\micro\m} (see \figref{fig:irdis-ref-spec}), so brighter areas generally indicate a higher ice abundance, and darker areas indicate lower ice abundance. The reference map (bottom) shows the visible light mosaic of Ganymede from \citet{usgs2013ganymede}. The maps show orthographic projections of Ganymede's northern and southern hemispheres and an equirectangular projection covering all of Ganymede's surface. \ang{0} is the sub-jovian longitude and \ang{180} is the anti-jovian longitude.
		\label{fig:spectral-ratio}}
\end{figure*}

The most dominant features of Ganymede's near-IR spectrum are the water ice absorption bands, particularly around \SI{1.5}{\micro\m} and \SI{2}{\micro\m} \cite{pilcher1972galilean,clark1980galilean,calvin1995spectra}. The depths of these bands can be used to provide a qualitative indication of water ice distribution, where stronger absorption implies a higher water ice abundance \cite{hapke1993book}. Ganymede's crust is thought to mainly consist of water ice \cite{schubert2004interior}, so the water ice distribution also shows the spatial distribution of non-ice contaminants on Ganymede's surface.

\figref{fig:spectral-ratio} shows the strength the \SI{2}{\micro m} water ice band, as measured by the SPHERE IRDIS instrument. The absolute depth of the \SI{2}{\micro\m} absorption varies significantly with the water ice grain size, however the relative depth (i.e. $2.25/\SI{2.11}{\micro\m}$) is similar for different grain sizes (see \figref{fig:irdis-ref-spec}). Therefore, we use the $2.25/\SI{2.11}{\micro\m}$ spectral ratio as the majority of the variations in this relative band depth are caused by variations in water ice abundance rather than varying ice properties.

The dominant trend in \figref{fig:spectral-ratio} is higher water ice abundances in Ganymede's younger bright terrain and low abundances in older dark terrain. The boundaries between the dark regiones and bright sulci are well defined showing the clear contrast between the two types of terrain. Relatively small scale features are visible, such as Nineveh Sulcus cutting through Perrine Regio (\ang{25}N, \ang{55}W) and the pattern of Sulci in and around Barnard and Nicholson Regio (\ang{\sim10}S, \ang{\sim0}W). There is also a notable increase in ice abundance on Ganymede's polar caps at latitudes above \ang{\sim45}.

The strongest ratios, and therefore likely highest ice abundances, appear around impact craters, with Tros (\ang{11}N, \ang{27}W) and Osiris (\ang{38}S, \ang{166}W) craters particularly prominent in \figref{fig:spectral-ratio}. This pattern is consistent with Ganymede's icy crust gradually being contaminated by non-ice materials over time and impact events excavating and exposing young pristine ice \cite{johnson1983global,mccord1998non,ligier2019surface}. Similarly, Sulci are much younger than Ganymede's dark terrain \cite{pappalardo2004geology,johnson1983global}, explaining the higher relative ice abundance in Sulci.

\section{Spectral modelling} \label{sec:modelling}
Our mapped spectral cubes are analysed by fitting to laboratory spectra from reference cryogenic libraries. We use the Markov Chain Monte Carlo (MCMC) linear unmixing technique described fully in \citet{king2022compositional} to calculate fitted endmember abundance values for each observed spectrum.

This MCMC modelling technique models models the spectrum as a linear sum of endmembers. The modelled spectrum, $M$, is
\begin{equation}
	M(\lambda) = \sum_{i}w_i E_i(\lambda)
	\label{eq:linear-spectral-model}
\end{equation}
where $w_i$ are the abundances of the different endmembers $E_i$. These abundances are constrained to physically possible values, $0 \leq w_i \leq 1$, and must sum to one, $\sum_{i} w_i = 1$. MCMC modelling repeatedly simulates the observed spectrum and produces a posterior distribution for the endmember abundances \cite{bayes1763lii, foreman2013emcee}. The modelling routine uses the measured uncertainties in the observed spectra (see \secref{sec:reduction}), meaning that the modelled posterior distributions therefore account for degeneracies and correlations between spectral endmembers and uncertainties in the observations \cite{king2022compositional}. We can sample these posterior distributions to calculate the best estimate abundance values and their associated uncertainties. We use the median of the abundance distribution as the best estimate and the 16th and 84th percentiles of the distribution to calculate the 1-$\sigma$ uncertainty.

The posterior distributions of multiple different endmembers can also be summed to calculate the total abundance and associated uncertainty of a combination of specific endmembers. The combined posterior distribution accounts for any degeneracies and correlations between the individual posterior distributions of the summed endmembers, so is significantly more powerful than simply adding the individual best estimate abundances of the individual endmembers. This technique is particularly useful for comparing different `classes' of endmembers (i.e. all ices/acids/salts/darkening agent) where there can be large degeneracies between individual endmembers which cancel when the aggregate abundance is calculated (see \secref{sec:roi} for examples).

The use of linear unmixing is appropriate for low spatial resolution ($\gtrsim 100$~km) ground-based observations where the observed spectrum for each pixel is naturally a linear combination of spectra from different geological units \cite{king2022compositional,ligier2016vlt}. This method does not account for any non-linear scattering caused by intimate mixtures on Ganymede's surface, so may therefore be responsible for some small residual errors after fitting \cite{shirley2016europa}. Radiative transfer models could be used to account for non-linear scattering, but this would require optical constants for all endmembers, many of which have never been measured at appropriate cryogenic temperatures \cite{ligier2019surface}.

\subsection{Spectral library}
\begin{table}
	\caption{Spectral library. Selected endmembers are shown in \figref{fig:example-endmembers}. \label{tab:spectral-library}}
	\centering
	\begin{tabular}{lll}
		\hline
		\textbf{Endmember}                  & \textbf{Name}               & \textbf{Reference}            \\
		\hline
		\ce{H2O} (refractive indices)       & Water ice                   & \citet{grundy1998temperature} \\
		\hline
		\ce{H2SO4.6.5H2O}                   & Sulphuric acid              & \citet{carlson1999sulfuric}   \\
		\ce{H2SO4.8H2O} (\SI{5}{\micro m})  &                             & \citet{carlson1999sulfuric}   \\
		\ce{H2SO4.8H2O} (\SI{50}{\micro m}) &                             & \citet{carlson1999sulfuric}   \\
		\hline
		\ce{Mg(ClO3)2.6H2O}                 & Magnesium chlorate          & \citet{hanley2014reflectance} \\
		\hline
		\ce{Mg(ClO4)2.6H2O}                 & Magnesium perchlorate       & \citet{hanley2014reflectance} \\
		\hline
		\ce{MgCl2.2H2O}                     & Magnesium chloride          & \citet{hanley2014reflectance} \\
		\ce{MgCl2.4H2O}                     &                             & \citet{hanley2014reflectance} \\
		\ce{MgCl2.6H2O}                     &                             & \citet{hanley2014reflectance} \\
		\hline
		\ce{MgSO4.6H2O}                     & Magnesium sulphate          & \citet{dalton2012low}         \\
		\ce{MgSO4.7H2O}                     &                             & \citet{dalton2012low}         \\
		\ce{MgSO4 Brine}                    &                             & \citet{dalton2007linear}      \\
		\hline
		\ce{Na2SO4.10H2O}                   & Mirabilite                  & \citet{dalton2007linear}      \\
		\hline
		\ce{Na2Mg(SO4)2.4H2O}               & Sodium magnesium sulphate   & \citet{dalton2012low}         \\
		\hline
		Black                               & Synthetic 0\% reflectance   & N/A                           \\
		White                               & Synthetic 100\% reflectance & N/A                           \\
		\hline
	\end{tabular}
\end{table}

The endmembers that make up our spectral library are given in \tabref{tab:spectral-library} and are based on those used for Europa in \cite{king2022compositional}. Our reference endmembers include hydrated sulphuric acid \cite{carlson1999sulfuric} and a variety hydrated salts \cite{hanley2014reflectance, dalton2012low, dalton2007linear}. The sulphuric acid spectra \cite{carlson1999sulfuric} do not cover wavelengths below \SI{1}{\micro\m}, so this limits our modelling to cover the \SIrange{1}{2.5}{\micro\m} spectral range. The spectral library is ultimately limited to spectra which cover this near-IR wavelength range have been measured in appropriate cryogenic temperatures which are representative of Ganymede's surface conditions. The hydrated salts included in our spectral library were detected in previous high spectral resolution modelling of Ganymede's surface composition \cite{ligier2019surface}. Reference spectra were convolved to the appropriate wavelengths, spectral resolutions and filter widths for use in modelling the SPHERE and NIMS spectra.

As in \citet{king2022compositional}, the water ice reflectance spectra are calculated from measured refractive indices \cite{grundy1998temperature} using the Hapke bidirectional reflectance model \cite{hapke1993book}. The modelled spectra were simulated using a phase angle of \ang{30} so their photometric properties were consistent with laboratory measured spectra \cite{dalton2012low} and the reduced datasets. Modelling was performed using a variety of simulated ice grain sizes between \SI{1}{\micro m} and \SI{1}{cm}. Initial models did not identify extremely small (\SI{<10}{\micro m}) or large (\SI{>3}{mm}) grains, so our final model presented here covers grains ranging \SI{10}{\micro m} to \SI{3}{mm}. The final modelled water ice spectra use five size bins ranging from \SI{10}{\micro m} to \SI{3}{mm} where the spectrum is an average of grain sizes within each bin \cite{king2022compositional}.

We also include two spectrally-flat synthetic spectra, one black (0\% reflectance) and one white (100\% reflectance). These two endmembers allow modelling of any featureless and spectrally-flat (i.e. grey) darkening agents present on Ganymede's surface which previous studies have found are necessary for the modelling of Ganymede's surface \cite{ligier2019surface}. The summed abundance of these two synthetic spectra can be used as an estimate for the abundance of the darkening agent, and the relative abundance of the black and white endmembers can provide an estimate for the reflectance of this material.

\section{Regions of interest} \label{sec:roi}
\begin{figure*}
	\centering
	\includegraphics[width=\linewidth]{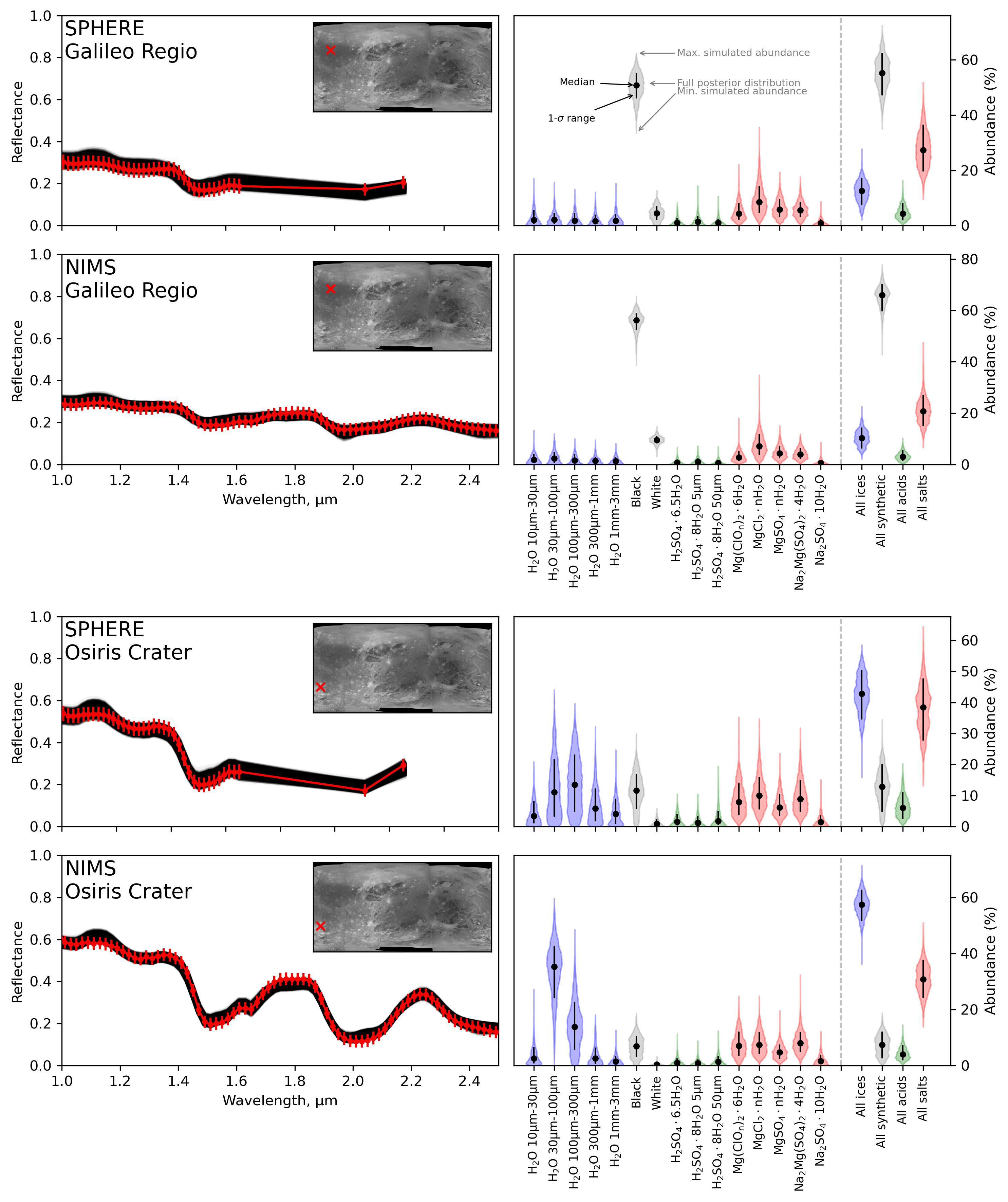}
	\caption{Fit result for spectra from Galileo Regio and Osiris Crater. The left hand column shows the observed spectrum (red) and the `family' of MCMC fitted spectra (black). The uncertainties on the observed spectra were calculated by measuring the noise level in the background sky pixels for each observation. The right hand column shows the violin plots for the fitted posterior abundance distributions where the black dots give the best estimate abundance, the black lines show the 1-$\sigma$ uncertainty and the coloured region shows the full posterior distribution.
		\label{fig:violin-both}}
\end{figure*}
\begin{figure*}
	\centering
	\includegraphics[width=\linewidth]{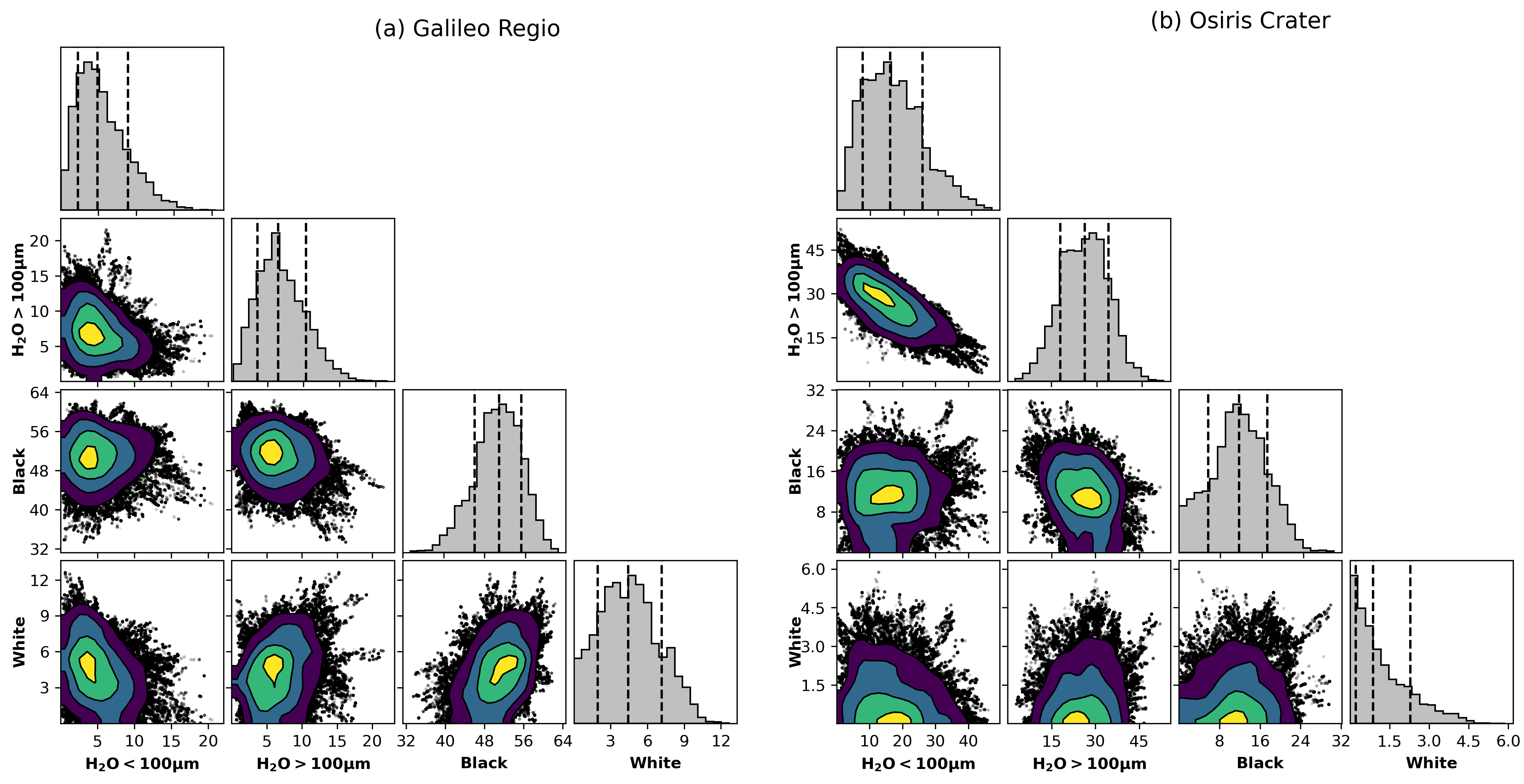}
	\caption{Corner plots showing relationships between percentage abundances for selected endmembers from SPHERE fits of Galileo Regio and Osiris crater. The shaded 2D-histograms give the posterior abundance distribution with brighter colours indicating a higher density. In low density regions towards the edge of the distribution, individual points of the distribution are plotted. Circular distributions imply no strong correlation between the abundance distributions of the two endmembers whereas skewed distributions imply a correlation between the abundances of those endmembers. The grey 1D-histograms show the posterior abundance distribution for the individual endmembers indicated by each column and the dashed lines show the 16th, 50th and 84th percentiles.
		\label{fig:corner-both}}
\end{figure*}

Initial modelling was performed on two regions of interest in Galileo Regio (\ang{145}W,~\ang{35}N) and Osiris Crater (\ang{166}W,~\ang{38}S). These two locations represent two geological extremes of Ganymede's surface, with the ancient dark terrain of Galileo Regio and the recent excavation of Osiris Crater. Modelling was performed for each location on our SPHERE dataset, and the co-located `G1GNGLOBAL01A' Galileo/NIMS dataset.

MCMC fit results for the two locations for both the SPHERE and NIMS datasets are summarised in \figref{fig:violin-both}. The two locations show contrasting compositions, with Galileo Regio dominated by the darkening agent and Osiris Crater predominantly water ice. Both locations have similarly low acid abundances and a mix of potential hydrated salts.

In addition to comparing the composition of the two locations, these regions of interest can be used to compare the capability of the ground-based SPHERE observations to the space-based NIMS observations. As shown previously for Europa \cite{king2022compositional}, both instruments show consistent results with similar best estimate values. Generally the uncertainties on the SPHERE fitted abundances are larger than the uncertainties for the corresponding NIMS fits. This difference is due to the larger uncertainties on the SPHERE dataset and the additional spectral range covered by NIMS providing more information to help lift degeneracies inherent in the narrower wavelength range of SPHERE.

The relationship between the posterior distributions of pairs of endmembers can be analysed using corner plots. \figref{fig:corner-both} shows the corner plots for selected endmembers for the Galileo Regio and Osiris crater regions of interest, where the shape of each 2D-histogram in a corner plot shows the relationship between the two endmember abundance distributions. Most pairs of endmembers have roughly circular distributions (e.g. between \ce{H2O} endmembers and the black endmember in \figref{fig:corner-both}) which implies that there is relatively little correlation between the individual endmember abundance distributions, so the abundances can be derived relatively independently. See Figure~10 in \citet{king2022compositional} for more corner plot modelling examples.

Water ice grain size endmembers, particularly in areas of high ice abundance such as Osiris crater (\figref{fig:corner-both}b) show a strong negative skew, implying negative correlation between the posterior distributions of different ice grain size endmembers. This is caused by spectral degeneracy between different ice grain size spectra making it difficult to constrain the ice grain size, even when the total ice abundance is well constrained (see \figref{fig:example-endmembers}). This effect can also be seen in \figref{fig:violin-both}  for the Osiris crater fits, where there are large uncertainties on the individual grain size endmembers, but these uncertainties cancel out leaving a much smaller uncertainty on the total water ice abundance.

Conversely, there is a positive skew between the synthetic black and white spectra in \figref{fig:corner-both}a. This suggests that the uncertainties are positively correlated such that finding a high abundance of one implies a high likelihood of finding a high abundance of the other. For this specific case, this correlation implies that the ratio of the black and white spectra (i.e. the reflectance of the darkening agent) is relatively certain, whereas there is a higher uncertainty in the total abundance of the darkening agent.

The trends shown in these regions of interest are representative of other locations on Ganymede. Therefore, ice grain size endmembers can be expected to be strongly negatively correlated, synthetic darkening agent endmembers positively correlated, and other endmembers relatively uncorrelated.

\section{Compositional maps}
\begin{figure*}
	\centering
	\includegraphics[width=\linewidth]{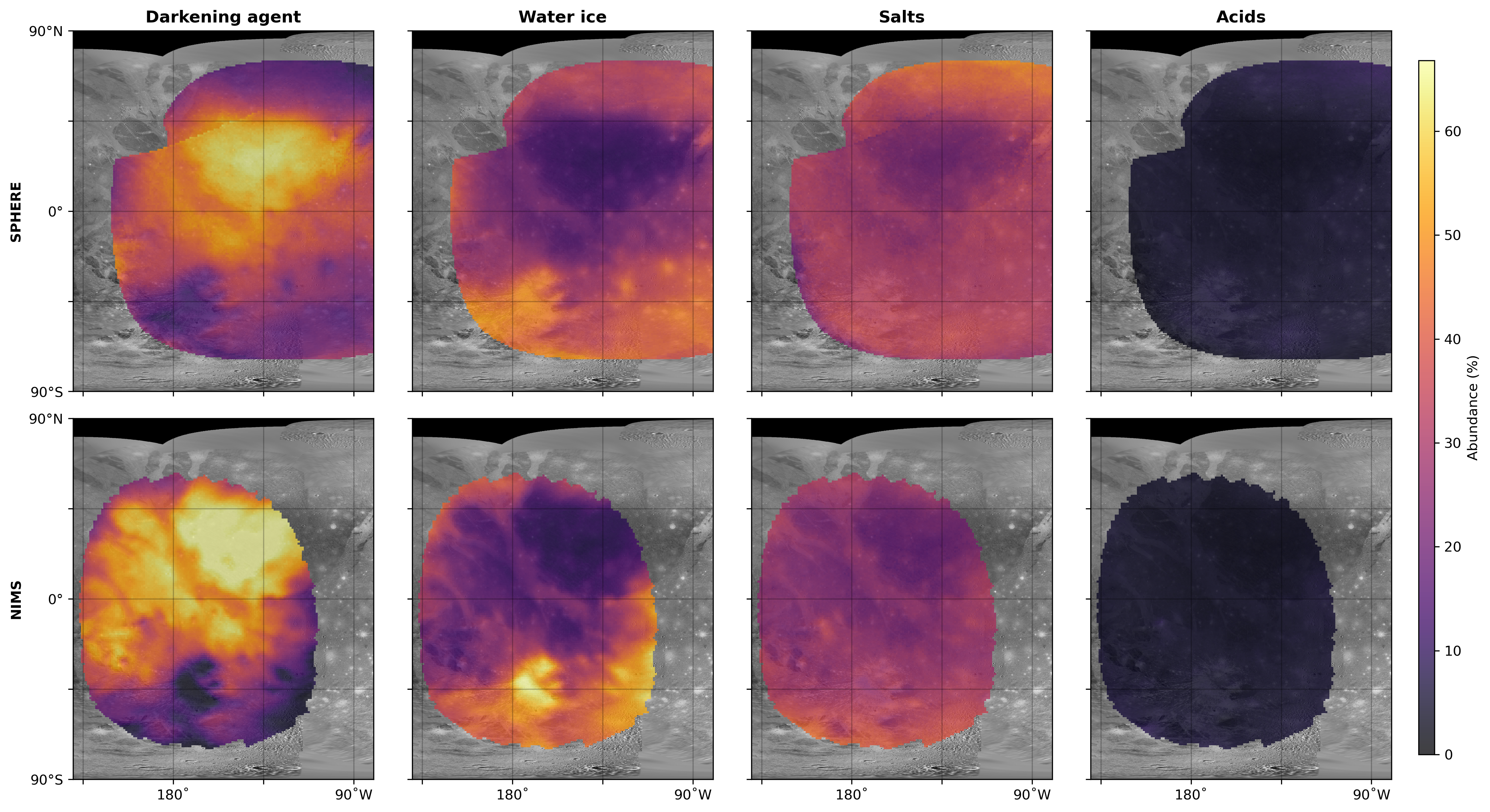}
	\caption{Best estimate modelled abundances for VLT/SPHERE (top) and Galileo/NIMS (bottom) observations in the region they overlap. The values shown here are the best estimates that are calculated as the median of the posterior abundance distribution for each location.
		\label{fig:nims-comparison}}
\end{figure*}

Modelling was performed for each observed location from our SPHERE dataset using our MCMC model. This produces a posterior abundance distribution for each endmember (and class of endmembers) that is then sampled to produce maps of the best estimate abundances for different species.

\figref{fig:nims-comparison} shows the best estimate abundances for the different classes of endmembers (darkening agent, ices, salts, and acids) for the SPHERE and NIMS observations. As with the regions of interest in \figref{fig:violin-both}, the two datasets show consistent results with a surface dominated by the darkening agent in old terrain and water ice in younger terrain. Salts appear relatively uniformly distributed with slightly higher abundances in younger terrain, and acids appear to have much lower abundances. The similarity between the NIMS and SPHERE results in \figref{fig:violin-both} and \figref{fig:nims-comparison} gives us confidence in using the broader SPHERE spatial coverage to extend the surface composition maps beyond what was possible with Galileo. The modelling results using the near global coverage of our SPHERE dataset are discussed in more detail in the following sections.

\subsection{Darkening agent}
\begin{figure*}
	\centering
	\includegraphics[width=\linewidth]{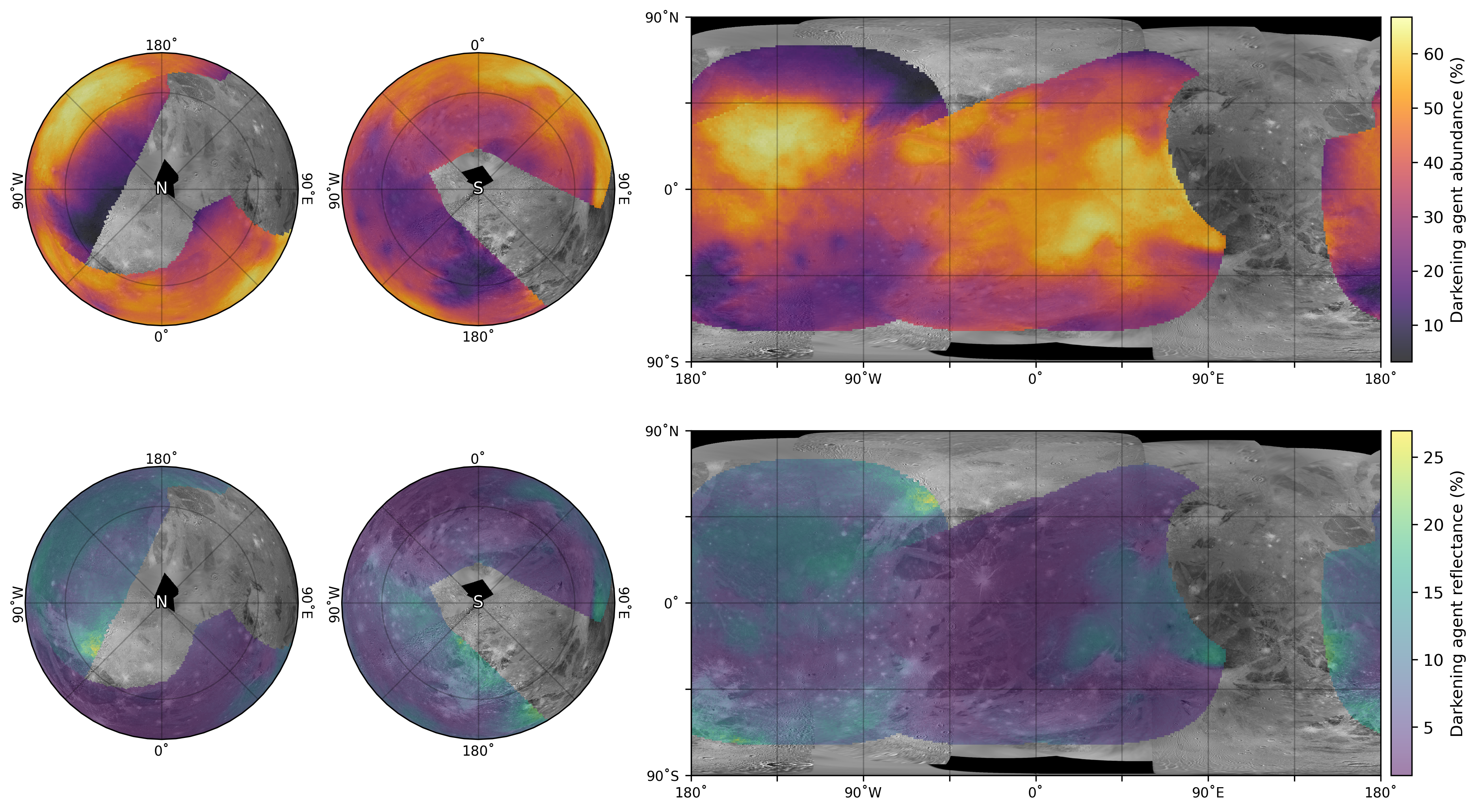}
	\caption{Darkening agent best estimate spatial distribution. The abundance of the darkening agent (top) is calculated as the sum of the abundances of the black and white synthetic spectra. The reflectance of the darkening agent (bottom) is calculated as the ratio of the white abundance to total darkening agent abundance. The uncertainty on the darkening agent abundance is $\pm\sim6$ percentage points.
		\label{fig:synthetic-map}}
\end{figure*}
\begin{figure}
	\centering
	\includegraphics[width=0.667\linewidth]{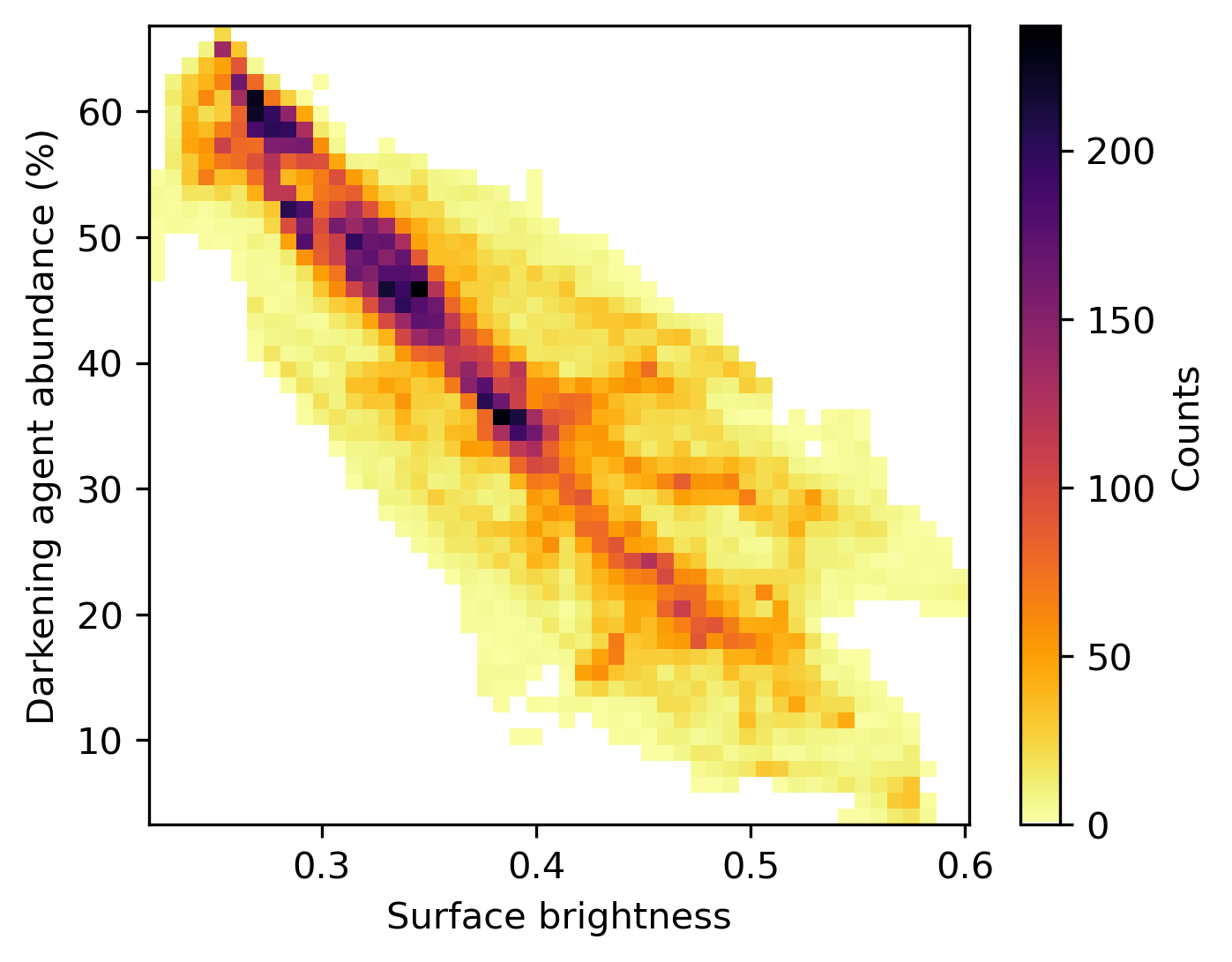}
	\caption{Correlation between darkening agent abundance and Ganymede's surface brightness. The darkening agent and brightness maps are projected to a regular 1~pixel-per-degree grid, then the pixel values are directly compared to produce this 2D histogram. Surface brightness is calculated using the visible light mosaic from \citet{usgs2013ganymede} which has been convolved with a 4~pixel wide gaussian to approximate the spatial resolution of the SPHERE observations.}
	\label{fig:synthetic-correlation}
\end{figure}

The modelling results shown in \figref{fig:synthetic-map} suggest that the spectrally-flat darkening agent is one of the key components of Ganymede's surface, with the expected high abundance in Ganymede's old dark terrain and lower abundance in the younger bright terrain. This old/young terrain contrast is consistent across all of Ganymede's surface, with Galileo Regio on the leading hemisphere and Melotte Regio on the trailing hemisphere showing the highest abundances of the darkening agent.

Across Ganymede's surface, there is a strong negative correlation between the darkening agent abundance and the visible surface brightness, as shown in \figref{fig:synthetic-correlation}. This suggests that the species responsible for darkening at visible wavelengths is also likely to be responsible for the darkening at near infrared wavelengths.

The implied reflectance of the darkening agent can be calculated by taking the ratio of the white synthetic abundance to the total synthetic abundance. The values shown in \figref{fig:synthetic-map} suggest a relatively uniformly low reflectance between 5\% and 20\% across the the majority of the observed area. The reflectance appears slightly higher in regions where the darkening agent abundance is higher, although the correlation is not perfect. The cause of this slight spatial variation is unclear, although it could be explained by small degeneracies with other modelled species, due to the low spectral resolution of the data, or some subtle compositional variations in the darkening agent material across the surface. This is discussed in more detail in \secref{sec:discussion}.

\subsection{Water ice}
\begin{figure*}
	\centering
	\includegraphics[width=\linewidth]{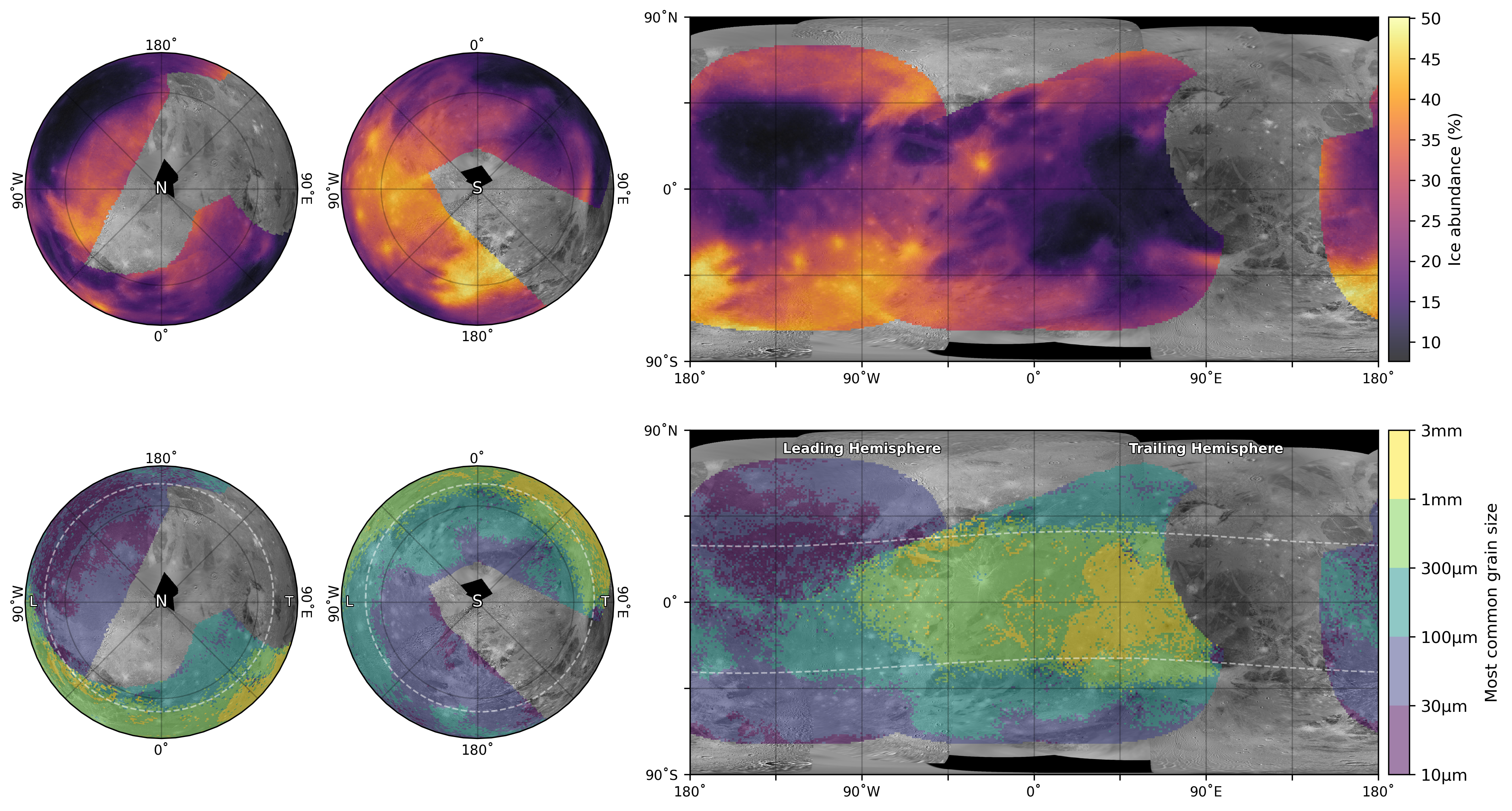}
	\caption{Water ice best estimate spatial distributions. The total water ice abundance (top) is highest in younger bright terrain, and particularly around impact craters. The uncertainty in water ice abundance varies between $\pm3$ percentage points in low abundance areas to $\pm7$ percentage points in high abundance areas. The grain size map (bottom) shows the ice size endmember with the highest individual abundance. Dashed lines indicate Ganymede's approximate open/closed field line boundaries from \citet{khurana2007origin}.
		\label{fig:ice-map}}
\end{figure*}
The water ice distribution in \figref{fig:ice-map} shows the same spatial pattern as the \SI{2}{\micro m} (\figref{fig:spectral-ratio}) absorption bands discussed previously, with high abundance in young bright terrain, and particularly around impact craters. There is also a generally higher ice abundance around Ganymede's polar caps (latitudes $\gtrsim\ang{45}$) which is particularly noticeable in the northern part of Galileo Regio. Water ice is highly anti-correlated with the darkening agent, with the combined ice and darkening agent abundance $\sim80\%$ when averaged over the observed area.

Ice grain sizes appear strongly correlated with latitude, with large (\SI{>100}{\micro m}) grains around the equator and smaller (\SI{<100}{\micro m}) grains towards the poles in \figref{fig:ice-map}. There is also a longitudinal trend in grain size, with the largest grains found at the equator on the trailing hemisphere, and smaller grains on the leading hemisphere, particularly around \ang{135}W to \ang{180}W. The distribution in grain sizes appears relatively uncorrelated with geological units (such as craters, sulci and regiones), suggesting that global scale trends are the dominant driver of variation in grain sizes. This water ice distribution is discussed in more detail in \secref{sec:discussion}.

\subsection{Hydrated sulphuric acid}
\begin{figure*}
	\centering
	\includegraphics[width=\linewidth]{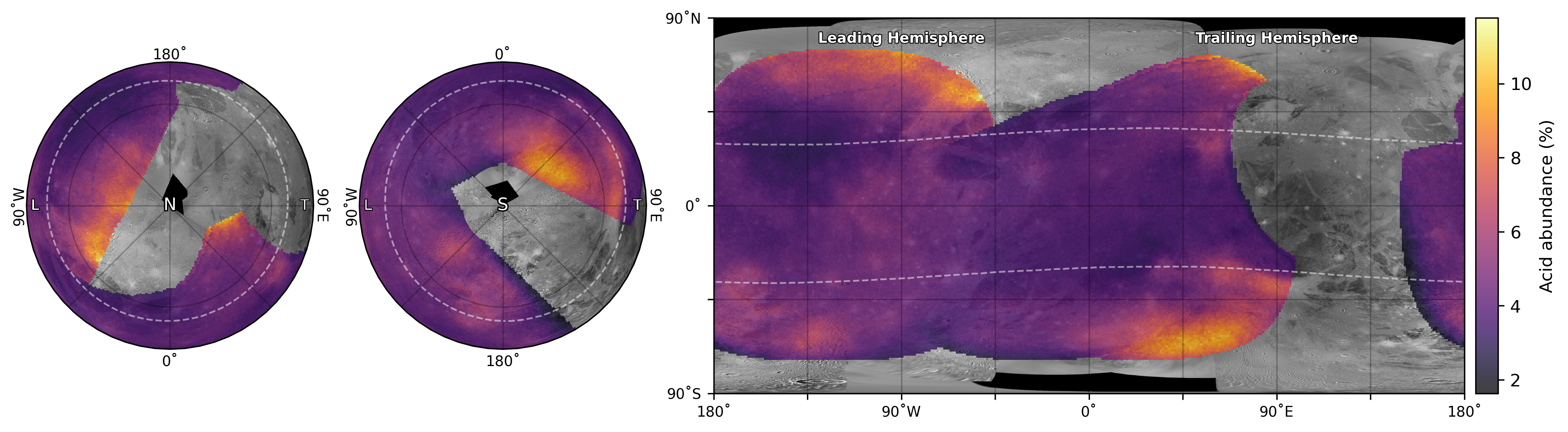}
	\caption{Hydrated sulphuric acid best estimate abundances. The 1-$\sigma$ uncertainty on acid abundance varies between $\pm2$ percentage points in low abundance areas to $\pm5$ percentage points in high abundance areas. Note that the magnitude of the spatial variations is generally less than the 2-$\sigma$ uncertainty on the abundances, so a uniform distribution of acid would be consistent with the data. Dashed lines indicate Ganymede's approximate open/closed field line boundaries from \citet{khurana2007origin}. The increased acid abundance at high latitudes on the leading hemisphere is consistent with the modelled spatial distribution of exogenic plasma bombardment \cite{liuzzo2020variability}.
		\label{fig:acid-map}}
\end{figure*}
Hydrated sulphuric acid abundances are shown in \figref{fig:acid-map} where the best estimate abundance is $<15\%$ across the entire observed area. The dominant trend in acid abundance is a higher acid abundance at high latitudes, particularly in the open magnetic field line regions at Ganymede's poles. The magnitude of the spatial variations in acid abundance is less than the 2-$\sigma$ uncertainty on the abundances, so this spatial variation cannot be detected with high confidence with SPHERE alone. However, the distribution is consistent with previous high spectral resolution modelling of Ganymede's acid distribution \cite{ligier2019surface} and hypothesised acid locations (see \secref{sec:discussion}).

The spatial distribution of acid at non-polar latitudes appears relatively uniform across most of the observed area, with very slightly higher abundance in bright terrain. However, these bright/dark terrain contrasts are very small and less than the uncertainties on the acid abundances, so may partly be a result of the slight degeneracy between hydrated acid and water ice spectra.

\subsection{Hydrated salts}
\begin{figure*}
	\centering
	\includegraphics[width=\linewidth]{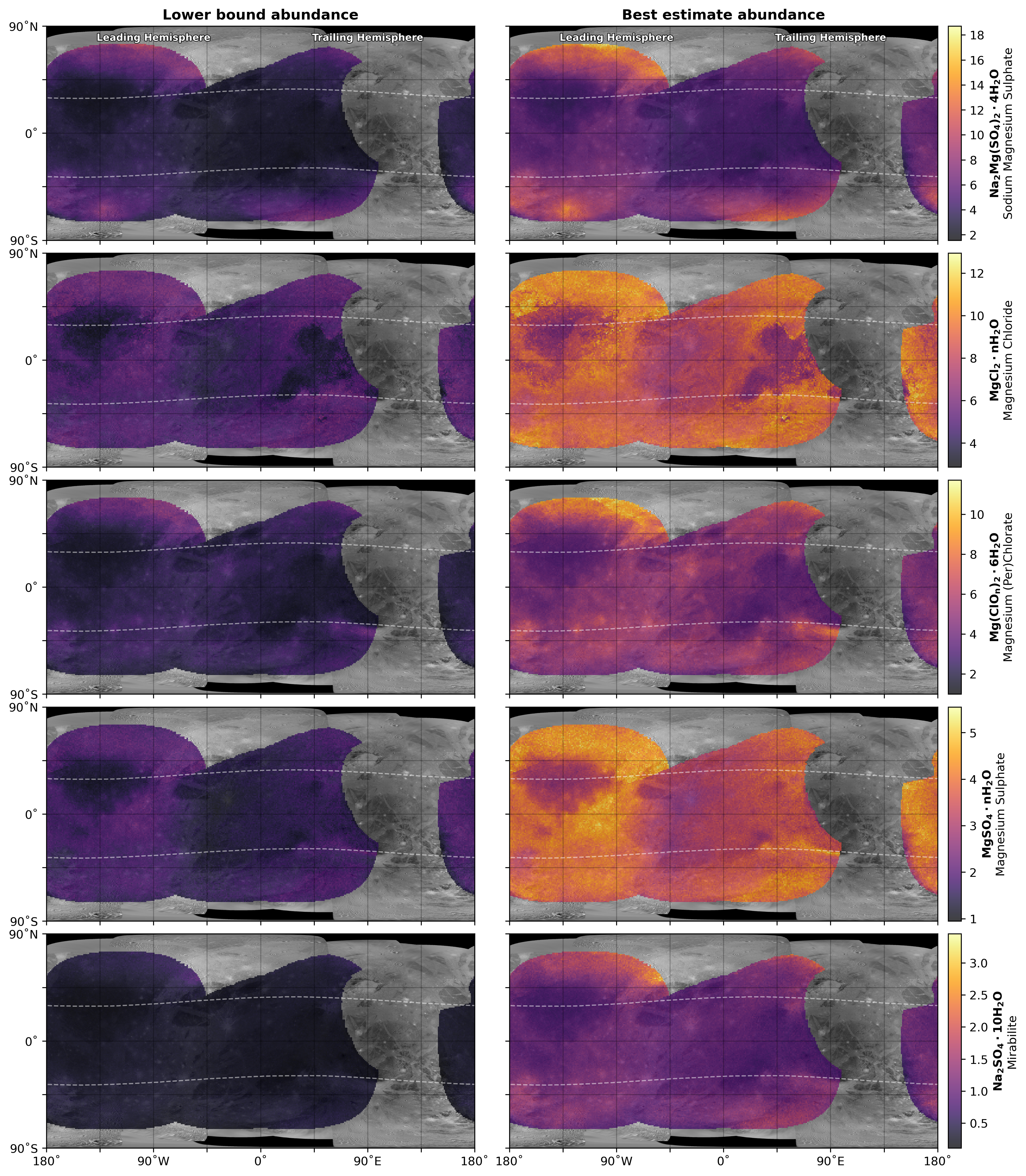}
	\caption{Modelled hydrated salt 1-$\sigma$ lower bound abundances (left) and best estimate abundances (right). A different abundance colourbar scale is used for each salt species to emphasise the variability of each species. Note that the magnitude of the spatial variations is generally less than the uncertainty on the abundances, so a range of spatial distributions would be consistent with the data. Dashed lines indicate Ganymede's approximate open/closed field line boundaries from \citet{khurana2007origin}.
		\label{fig:salt-map}}
\end{figure*}
The modelled lower bound and best estimate abundances for six different salt candidates are shown in \figref{fig:salt-map}. Although the averaged salt abundance was $\sim25^{+12}_{-8}\%$ over the observed area, there were no strong detections of any individual salt, with 1-$\sigma$ uncertainty lower bound on the retrieved abundances generally close to 0\% for the different salt candidates. Therefore, the observed spectra are consistent with each individual salt only existing in small quantities (i.e. a few percent abundance) or being completely absent. However, the overall salt abundance suggests that some combination of salts is likely to be present, even if the specific combination cannot be detected with confidence.

Three of the salt families: sodium magnesium sulphate (\ce{Na2Mg(SO4)2.4H2O}), magnesium (per)chlorate (\ce{Mg(ClO3)2.6H2O} \& (\ce{Mg(ClO4)2.6H2O}), and mirabilite (\ce{Na2SO4.10H2O}) show a similar spatial distribution to sulphuric acid (\figref{fig:acid-map}) with high abundances in the polar open field line regions. Sodium magnesium sulphate has the highest abundances, with abundances of $\sim15^{+17}_{-12}\%$ around the northern polar cap. Magnesium chlorate has a maximum abundance of $\sim7^{+7}_{-5}\%$ while magnesium perchlorate and mirabilite both have maximum abundances of $\sim3^{+4}_{-2}\%$.

Two other salt families, magnesium chloride (\ce{MgCl2.nH2O}) and magnesium sulphate (\ce{MgSO4.nH2O}) show much less latitudinal variation and appear to be generally correlated with younger bright terrain. Magnesium chloride has a maximum abundance of $\sim11^{+15}_{-8}\%$ and magnesium sulphate has a maximum abundance of $\sim4^{+6}_{-3}\%$. The relatively uniform spatial distribution of magnesium chloride and magnesium sulphate mean that they have abundances close to their maximum abundance across much of Ganymede's surface.

It should be noted that the spatial contrasts in abundances are less than the uncertainties on the salt abundances. Therefore, although these distributions represent the best estimate of the salt spatial distributions, there is a wide range of plausible spatial distributions that are consistent with the observed spectra.

\section{Discussion} \label{sec:discussion}
\begin{figure*}
	\centering
	\includegraphics[width=\linewidth]{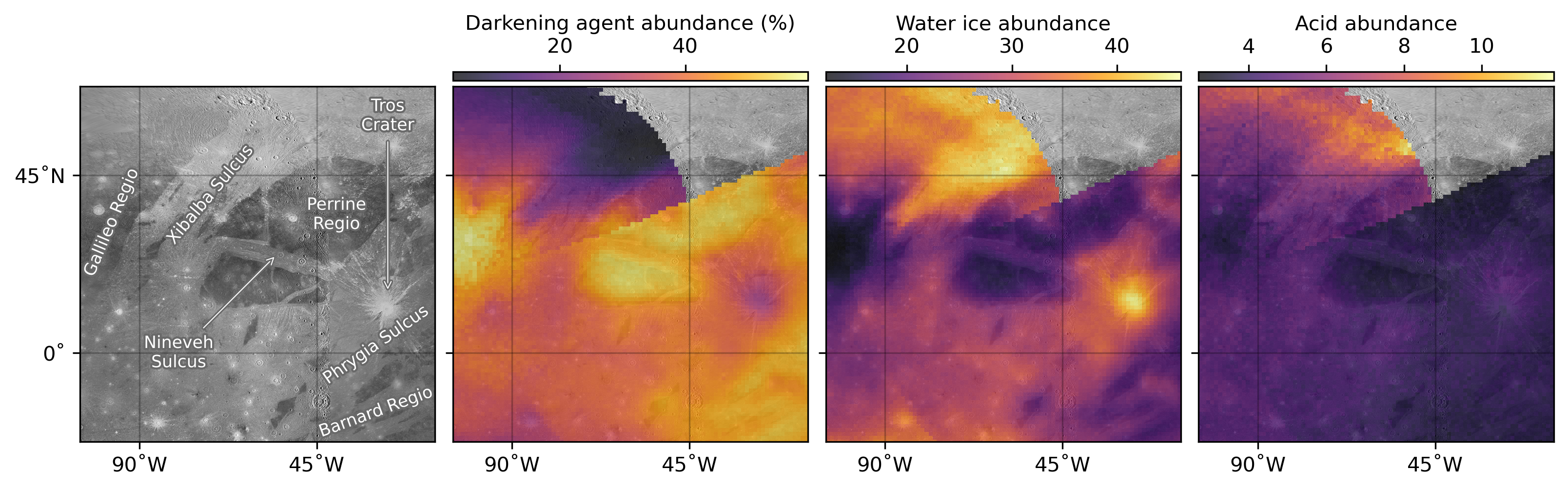}
	\caption{Best estimate modelled abundances in and around Perrine Regio, featuring Nineveh Sulcus and Tros Crater as examples of the compositional contrasts revealed by SPHERE.
		\label{fig:zoomed-map}}
\end{figure*}
The dominant compositional contrast on Ganymede's surface is the strong contrast between the old dark terrains and young bright terrains, which are mainly composed of the darkening agent and water ice, respectively. This is especially clear in the region around Perrine Regio, as shown in \figref{fig:zoomed-map}, where there are a range of major geological features (regiones, sulci and impact craters) in the same area. Note that this region was not observed by Galileo/NIMS with the full SPHERE spectral range (see \figref{fig:nims-resolution}). Water ice in particular appears well constrained to younger areas, with sharp changes in abundance around impact craters (e.g. Tros in \figref{fig:zoomed-map}) and between areas such as Xibalba Sulcus and Galileo Regio. This is consistent with Ganymede's surface consisting of an ancient icy crust covered with a layer of non-ice material and more recent tectonic features consisting of cleaner, more pristine ice, such as Nineveh Sulcus cutting across Perrine Regio in \figref{fig:zoomed-map} \cite{pappalardo2004geology,mccord1998non}.

It is notable how Tros impact crater and its associated ejecta in \figref{fig:zoomed-map} have an even higher ice abundance and lower darkening agent abundance than the surrounding Phrygia Sulcus. This is consistent with Ganymede's surface gradually being contaminated over time, with the very highest ice abundances in the young, pristine, impact craters contrasting with the older tectonic sulci. The sulci then in turn contrast with the ancient regiones, which have the highest darkening agent abundance and lowest water ice abundance. The water ice abundances, generally varying between 15\% and 50\% with higher abundances around the polar cap and lower abundances on the trailing hemisphere, are consistent with the abundances calculated from Galileo/NIMS \cite{mccord1998non,mccord2001hydrated} and VLT/SINFONI \cite{ligier2019surface}.

The darkening agent had a low albedo that generally varied between 5\% and 20\% across the observed area (\figref{fig:synthetic-map}). This albedo is slightly lower than the 24\% found by \citet{ligier2019surface}, however this previous study used a single reflectance value at all locations calculated from observations with slightly different spatial coverage. Therefore it is not directly comparable to our values, which are calculated independently for each observed location. The exact composition of this spectrally-flat darkening agent is unclear, although hydrated silicates, carbonaceous compounds, or potentially some hydrated salts are the most likely candidates \cite{ligier2019surface,pappalardo2004geology}. The lack of unique spectral features makes any determination of the composition of this dark material very difficult, so observations with much higher spectral resolution, covering different spectral ranges, or using different techniques are likely to be crucial to constrain the composition of this darkening agent.

The water ice grain size in \figref{fig:ice-map} varies gradually at a global scale. There are no rapid changes in grain size around geological units (e.g. impact craters, Regio/Sulci boundaries), suggesting that the ice grain size is relatively unaffected by local endogenic processes. The variations in grain size can therefore be attributed to latitudinal and longitudinal effects driven by Ganymede's thermal and radiation environments. The range of ice grain sizes (generally \SI{30}{\micro\m} to \SI{1}{mm}) agrees well with previous studies \cite{mccord2001hydrated,ligier2019surface,stephan2020h2o}, and our observations allow the near-global mapping of the ice grain size at \SI{\sim100}{km} scale resolution.

The latitudinal trend in ice grain size, with larger grains nearer the equator and smaller grains towards the poles, is consistent with the trend previously identified in VLT/SINFONI \cite{ligier2019surface} and NIMS \cite{stephan2020h2o} observations. This latitudinal gradient is likely to be mainly caused by the varying thermal environment, with higher average equatorial temperatures creating an environment more favourable for grain merging, therefore encouraging larger grain sizes. The thermal gradient is also likely to be responsible for thermal migration of \ce{H2O} to higher latitudes \cite{stephan2020h2o}. At the relatively warm equator, small ice grain sizes will sublimate at a faster rate than large grains (due to their increased surface-area-to-volume ratio) while at the colder poles, the sublimation rate is significantly slower, leading to a latitudinal gradient in equilibrium grain sizes from the warm equator to the cold polar caps. Diurnal temperature variations are too rapid to cause any significant day-night grain size trends in the longitude direction \cite{clark1983frost}.

The longitudinal trend in water ice grain sizes can similarly be explained by varying plasma bombardment around Ganymede's low latitudes. The most intense radiation at equatorial latitudes occurs on the trailing hemisphere, in the same region where the largest grains are found \cite{liuzzo2020variability}. Similarly to thermal effects, more intense radiation driven sputtering preferentially destroys smaller ice grains due to their surface area to volume ratio, leaving a larger equilibrium grain size in more heavily bombard equatorial areas \cite{clark1983frost}.

Ganymede's open magnetic field lines at the poles mean that plasma bombardment, and therefore sputtering, is also more intense at high latitudes \cite{khurana2007origin}. If sputtering eroding small grains was the only process present, large grains would be found at the poles rather than the small grains which are observed \cite{clark1983frost}. However, sputtering under Ganymede's polar conditions may also lead to small ice grains being redeposited on the surface, reducing the average grain size \cite{johnson1997polar,khurana2007origin,ligier2019surface}. Therefore, the observed latitudinal grain size trend is likely to be a combination of the strong thermal gradient and sputtering driven redistribution of small grains.

The apparent increased acid abundance at high latitudes (\figref{fig:acid-map}) is consistent with increased jovian magnetospheric plasma bombardment in Ganymede's polar open magnetic field line regions and a similar detection of sulphuric acid with VLT/SINFONI \cite{ligier2019surface}. This plasma bombardment delivers sulphur ions which can then produce sulphuric acid through radiolysis, ultimately leading to acid abundances correlated with exogenic plasma bombardment. This process is very prominent and has been well studied on Europa's trailing hemisphere \cite{carlson2005distribution, brown2013salts, ligier2016vlt, king2022compositional}, but is much more subtle on Ganymede due to the presence of its intrinsic magnetic field which shields Ganymede's equatorial latitudes from intense plasma bombardment \cite{liuzzo2020variability}. The MCMC derived uncertainties on the acid abundance makes it difficult to confidently detect acid with the SPHERE dataset alone, however the consistency between the observed distribution, previous detections \cite{ligier2019surface} and plasma bombardment locations increases the confidence in the detected acid distribution.

As previously demonstrated on Europa \cite{king2022compositional}, the lack of strong unique spectral features in hydrated salt spectra mean that individual salt species cannot be detected with confidence using these low spectral resolution near-IR spectra. Therefore, the degeneracies between the salt endmembers mean that the observed Ganymede spectra are consistent with a range of possible salt mixtures, and are consistent with the absence of individual salt species (see \figref{fig:salt-map}). Although this means a detailed analysis of hydrated salts and their origins is not possible, the higher spatial resolution offered by SPHERE can be used to infer potential spatial distributions that the salts may have if they are present. These potential spatial distributions can then be compared to other high spectral resolution observations which can more confidently detect species in a general area.

Sodium magnesium sulphate and magnesium chlorates appear to have a similar spatial distribution (\figref{fig:acid-map}) to that of acid, which hints that they may have similar origins. The correlation with regions of more intense plasma bombardment suggests that, if present, the salts may be related to exogenic processes. The delivery of ions such as \ce{S+} and \ce{Na+} from Jupiter's magnetospheric plasma could provide some of the ingredients necessary for the creation of the salts on Ganymede's surface \cite{carlson2009europa}.

Magnesium chloride and sulphate appear more uniformly distributed across Ganymede's surface with a slight correlation with younger bright terrain. This suggests a potential endogenic origin for these salts, if they are present, where they may be related to the distribution of fresh material (e.g. \ce{Mg^2+} or \ce{SO4^2-} ions) from Ganymede's subsurface ocean, which would have a higher abundance in younger terrain. \citet{ligier2019surface} identified a similar slight correlation of sulphate salts with younger terrain, suggesting that this magnesium sulphate distribution may be related to endogenic processes on Ganymede.

As demonstrated by the uncertainties in the modelled salt abundances produced by the MCMC routine, higher spectral resolution, spatial resolution and broader wavelength coverage will be crucial to positively identify and constrain the presence of individual salts on Ganymede's surface. Nevertheless, the combination of our MCMC spectral modelling and the high-spatial resolution of SPHERE allows us to produce global maps of Ganymede's reflectivity and surface composition. We hope that these global high resolution maps can provide useful context for future close-up views at regional scales.

Future observations with new generations of instruments will be able to more precise studies into Ganymede's surface composition. Juno will be able to use JIRAM's high spectral resolution and the unprecedented spatial resolution offered by Juno's position in the Jupiter system to study and constrain Ganymede's non-ice surface composition at small scales \cite{mishra2021bayesian}. New ground and space based telescopes such as ELT/HARMONI and JWST/NIRSPEC will enable spectroscopy to identify narrow features whilst also acquiring high spatial resolution global scale mapping to identify small-scale compositional contrasts.

\section{Conclusion} \label{sec:conclusion}
The near-infrared spectral observations with SPHERE have enabled compositional mapping covering the majority of Ganymede's surface at high spatial resolution (\SI{\sim100}{km}). Modelling of co-located SPHERE and Galileo/NIMS observations showed the two instruments produced consistent estimates of abundances of ices, acids, salts and synthetic darkening agents and validated the utility of SPHERE's low spectral resolution and high spatial resolution observations.

The use of our MCMC modelling technique allows us to calculate full posterior abundance distributions for each modelled endmember. This enables the more detailed investigation and understanding of the uncertainties on fitted abundance values when compared to simple linear fitting. As on Europa \cite{king2022compositional}, these uncertainties are particularly useful where they show that confident detection of individual salt species is not possible with the spectral resolution and coverage of SPHERE and NIMS, due to the significant degeneracies between hydrated salt spectra.

The identified spatial distributions of species appears consistent with previous Ganymede studies, and the high spatial resolution and near-global coverage of SPHERE has enabled us to accurately spatially constrain features across the majority of Ganymede's surface. High water ice abundance was identified in young areas, such as bright sulci and particularly impact craters, while older dark terrain was dominated by a low albedo darkening agent. The water ice grain size distribution variation appeared to be primarily driven by thermal gradients latitudinally and radiation variation longitudinally. Higher abundances of hydrated sulphuric acid, sodium magnesium sulphate and magnesium chlorates were very tentatively found in polar regions where the surface is unprotected by Ganymede's magnetic field, suggesting a potential endogenic origin for these species if they are present. The best estimate abundances for magnesium chloride and magnesium sulphate salts appeared more uniformly distributed, with a slight correlation with younger terrain, suggesting that, if present, these salts may be endogenous.

Future studies will help to further constrain the surface composition of Ganymede by expanding the spectral coverage and resolution of near-infrared observations. Other spectral ranges and higher spectral resolutions may be able to identify signatures of specific species that are not visible at the relatively low spectral resolutions used in this study, and greater spatial coverage will help to further constrain compositional correlations with geological units. New cryogenic laboratory measurements will also help to expand the coverage of spectral libraries, and enable re-analysis of existing observations with increasingly detailed and wide-ranging reference spectra.

The icy Galilean moons (Europa, Ganymede and Callisto) are due to be studied in the coming decade by ESA's \textit{Jupiter Icy Moons Explorer} (JUICE) and NASA's \textit{Europa Clipper}. JUICE in particular will ultimately orbit Ganymede, providing detailed global data of the entire moon \cite{grasset2013jupiter}. This planned exploration of the Jovian system with JUICE and Europa Clipper will help to significantly increase the amount of high spatial resolution observations of Ganymede, ultimately helping to constrain compositional variations to scales \SI{<1}{km}, improving our understanding of the local variation of Ganymede's surface composition.

\appendix
\section{Spectral library} \label{app:example-endmembers}
\begin{figure*}
	\centering
	\includegraphics[width=\linewidth]{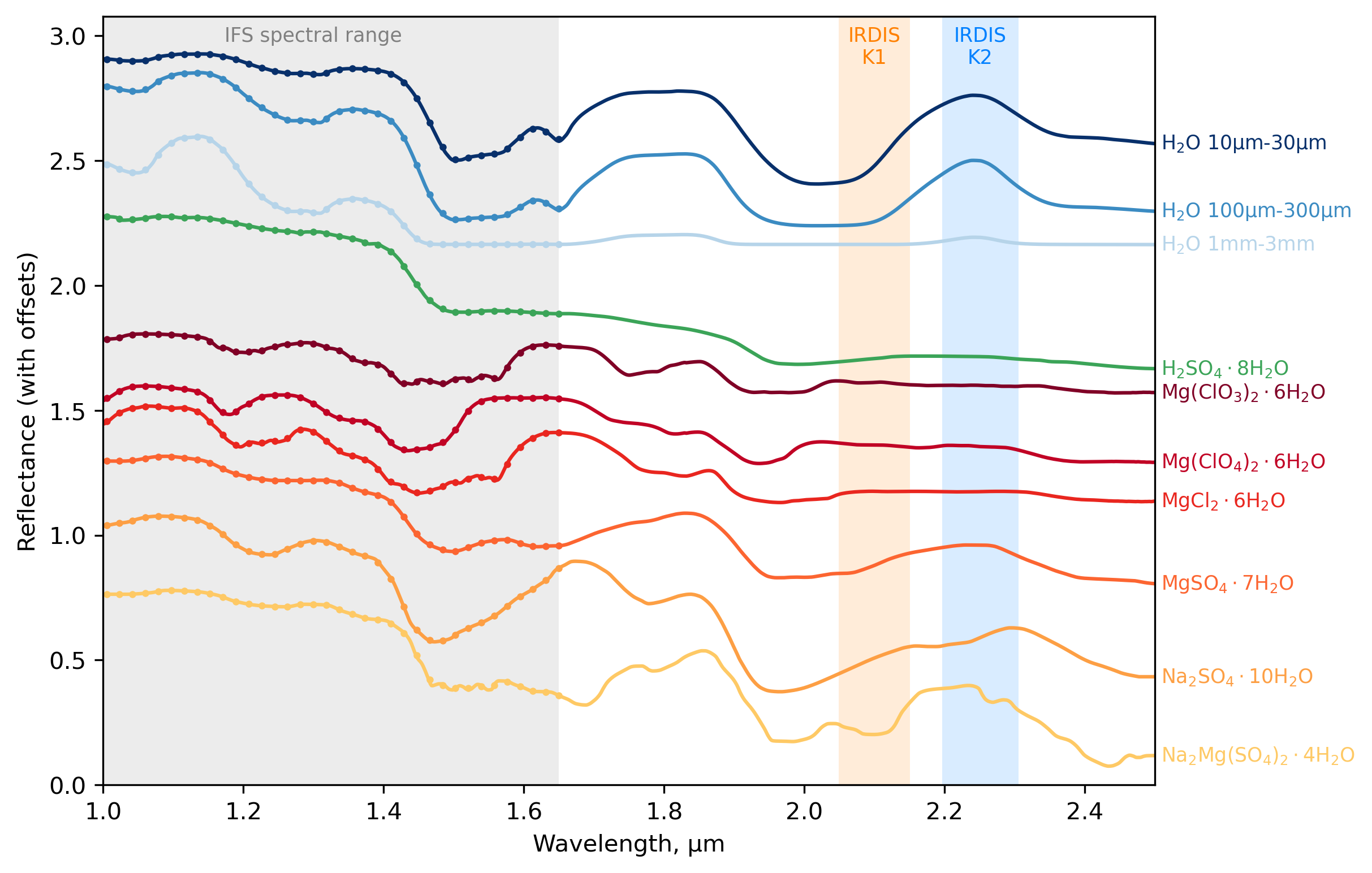}
	\caption{Example endmembers used for spectral modelling. The shaded areas show the spectral range covered by the SPHERE IFS instrument and the two IRDIS filters, and the dots show the reflectance in the IFS spectral bins. The NIMS spectra used in this study cover the whole spectral range shown in this figure. Spectra are vertically offset for clarity.
		\label{fig:example-endmembers}}
\end{figure*}

\figref{fig:example-endmembers} shows selected endmembers from our spectral library.

\section{Open Research}
The datasets used in this study can be found at \url{https://doi.org/10.5281/zenodo.6390443} \cite{king2022zenodo_ganymede}. This includes the reduced VLT/SPHERE and Galileo/NIMS reflectance cubes, and MCMC model outputs for each observation which contain derived endmember abundance values and associated uncertainties.

\acknowledgments
Based on observations collected at the European Organisation for Astronomical Research in the Southern Hemisphere under ESO programmes 60.A-9372(A) and 105.20EC.001. Oliver King was supported by a Royal Society studentship grant at the University of Leicester. Leigh Fletcher was supported by a Royal Society Research Fellowship and European Research Council Consolidator Grant (under the European Union's Horizon 2020 research and innovation programme, grant agreement No 723890) at the University of Leicester. This research used the ALICE High Performance Computing Facility at the University of Leicester. We are extremely grateful to Fraser Clarke, Niranjan Thatte and Nicolas Ligier for their support in the initial analysis of the science verification observations of Ganymede in 2015.

\bibliography{references}
\end{document}